\pdfoutput=1
\documentclass{JINST}

\title{Operation of the T2K time projection chambers}

\author{Georges Vasseur
\thanks{Proceedings of MPGD2011, the $2^{nd}$ International Conference on 
Micro Pattern Gaseous Detectors, 
August 29 - September 1, 2011, Kobe, Japan}\\
  CEA, IRFU, SPP, Centre de Saclay, \\
  F-91191 Gif-sur-Yvette, France\\
  E-mail: \email{georges.vasseur@cea.fr}}

\abstract{The three time projection chambers of the T2K near detector are 
micro pattern gaseous detectors based on bulk micromegas technology. 
They have been operated successfully during the first two physics runs 
of the experiment.
Their design, operation, and performance are presented.}


\begin{document}

\section{The T2K experiment and ND280 near detector}

The Tokai to Kamioka (T2K) experiment~\cite{t2k} is 
a long baseline neutrino oscillation experiment taking place in Japan. 
A 30 GeV proton accelerator at the J-PARC facility in Tokai is used to produce 
a high intensity $\nu_\mu$ beam with a peak energy around 0.7 GeV that is sent 
to a near detector in Tokai and to the SuperKamiokande water Cerenkov far
detector, 295 km away. 
The main two goals of the experiment are to get a first measurement of 
the mixing angle $\theta_{13}$ through $\nu_e$ appearance and  to measure 
precisely the atmospheric parameters $\theta_{23}$ and $\Delta m^2_{32}$ 
through $\nu_\mu$ disappearance.

\begin{figure}[htb]
\begin{center}
\includegraphics[width=.6\textwidth]{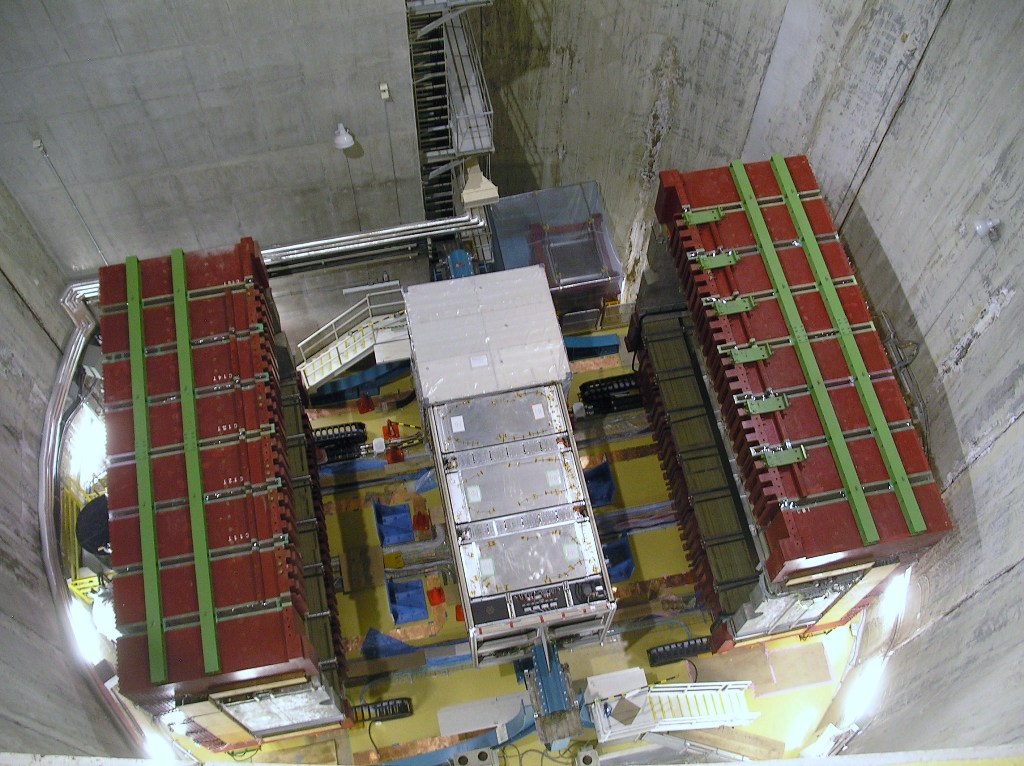}
\end{center}
\caption{View of ND280 from the top of the pit with the magnet open. 
The three TPCs can be seen sandwiching the two FGDs between the P0D (top) 
and the downstream ECAL (bottom).}
\label{fig:topview}
\end{figure}

The off-axis near detector ND280, shown in figure~\ref{fig:topview}, 
is located 280 m downstream of the production target. 
It consists of several detectors inside the UA1 magnet, 
producing a horizontal magnetic field of 0.188 T. 
The tracker is composed of two scintillator-based fine grained detectors (FGD) 
acting as active neutrino target interleaved 
with three time projection chambers (TPC). 
Upstream is a $\pi^0$ detector (P0D), 
all around are electromagnetic calorimeters (ECAL) and
inserted in the magnet yoke are muon range detectors. 
The goal of the near detector is to characterize the neutrino beam 
before oscillation: in particular to measure the $\nu_\mu$ energy spectrum, 
to determine the $\nu_e$ intrinsic contamination in the beam, 
and to study other background processes.

\section{The time projection chamber design}

The TPCs~\cite{tpc} are used for tracking and particle identification. 
A relatively modest resolution of 10\% on the transverse momentum measurement 
at 1 GeV is enough as the neutrino energy estimation is limited 
by the Fermi motion of the stuck nucleon at the level of about 10\%. 
The overall momentum scale however needs to be known at better than 2\% 
for $\Delta m^2_{32}$ measurement. 
Finally, to be able to separate electrons from muons and measure 
the $\nu_e$ contamination in the beam, 
a resolution on the energy loss dE/dx better than 10\% is needed.

\begin{figure}[htb]
\begin{center}
\includegraphics[width=.356\textwidth]{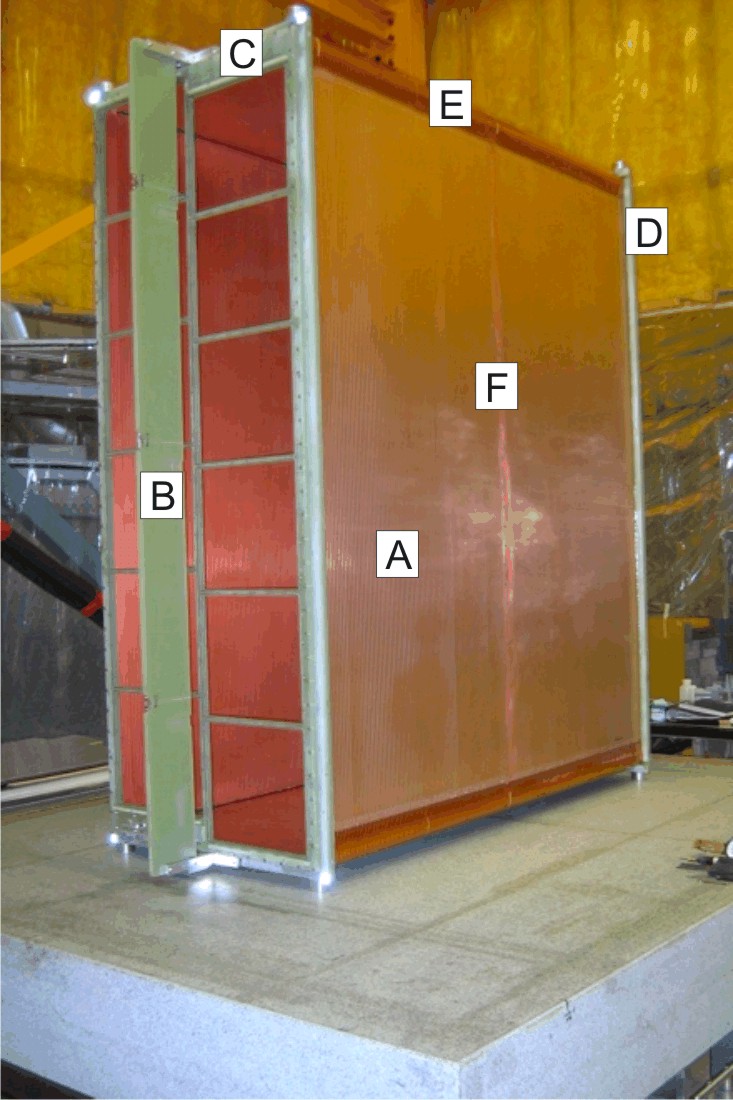}
\includegraphics[width=.4\textwidth]{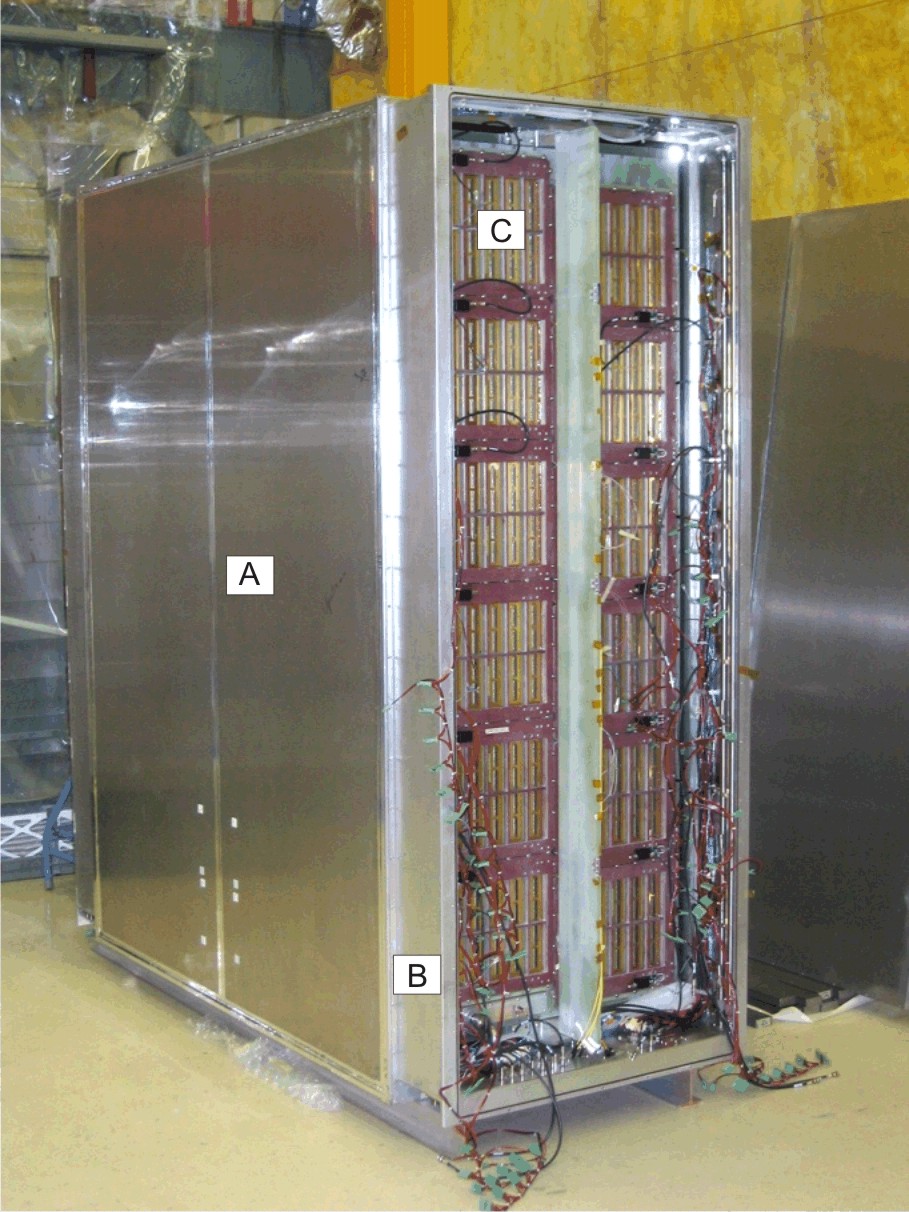}
\end{center}
\caption{Left: TPC inner box with (A) box wall, (B) module frame stiffening 
plate, (C) module frame, (D) endplate, (E) fiel-reducing corner, and (F)
central cathode. Right: outer box with (A) box wall, (B) service spacer, and 
(C) a micromegas module.}
\label{fig:tpc}
\end{figure}

A TPC has a rectangular shape, as shown in figure~\ref{fig:tpc}. 
Its tracking length along the beam direction is 72 cm and its active height is 
about 200 cm. 
The electrons drift along the electric field direction from the central cathode
to both endplates of the TPC,
each of which is instrumented with twelve micromegas (MM) modules,
in two columns of six. 
This gives a maximum drift distance of 90 cm on each side 
and a total active area for the three TPCs of 9 $\rm{m}^2$. 
T2K TPCs were indeed the first large size TPCs using micro pattern gaseous 
detectors. 

Mechanically, a TPC consists of two gastight boxes, one inside the other. 
The inner box, made of copper clad G10 laminated panels, is divided 
by the central cathode and has one module frame at each end to support 
the MM modules. 
The electric field is shaped by copper strips. 
Requirements are set on the cathode flatness and module plane planarity 
at the 0.1 and 0.2 mm level to minimize electric field distortions. 
The outer box is made of aluminum/rohacell laminated panels. 
Spacers behind the detectors are used for services: cooling, 
readout electronics, voltage connection, temperature probes, 
and calibration system.

Both of the boxes are filled with gas. 
The gas mixture in the inner drift volume is based on argon at 95\% and 
contains also 3\% of freon and 2\% of isobutane. 
This gives a sufficiently high gain, a large drift velocity of 7.5 cm/$\mu$s, 
and a low transverse diffusion coefficient of about 250 $\mu$m/$\sqrt{\rm cm}$
with the magnetic field on.
The gas is circulating and two small TPCs are used to monitor 
the supply and return gas, in particular to measure the gain and 
drift velocity. 
As for the outer volume it is filled with CO2 for electrical insulation.

The T2K TPCs are based on bulk micromegas technology~\cite{mm}. 
In the drift region between the cathode set at -25 kV and the mesh 
set at -350 V, with an electric field of 280 V/cm, 
the electrons created by the ionization of the gaz 
when a charged particle goes through the TPC drift to the mesh. 
Then they enter the 128 $\mu$m thick amplification region between the mesh and 
the grounded pads, where the electric field is about 100 times higher, 
resulting in an avalanche. 
For a high voltage of 350 V between the mesh and the pads, 
the TPC has a gain of 1500.

\begin{figure}[htb]
\begin{center}
\includegraphics[width=.6\textwidth]{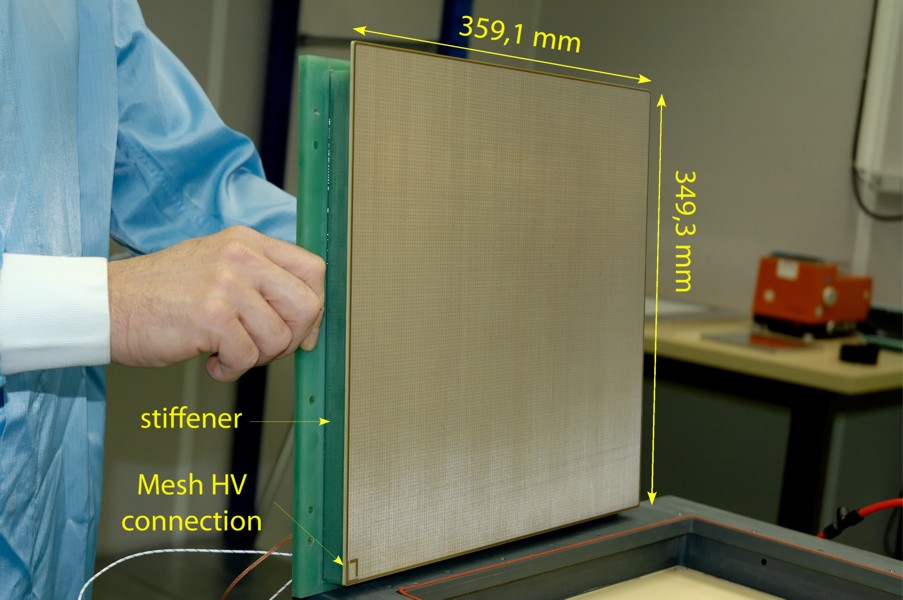}
\end{center}
\caption{View of a micromegas module.}
\label{fig:mm}
\end{figure}

Figure~\ref{fig:mm} shows one MM module. 
It is $36 \times 35 \, \rm{cm}^2$ and has 1726 active pads, 
$6.9 \times 9.7 \, \rm{mm}^2$. 
As there are 12 modules on each TPC endplate, we have in total 72 modules 
for the 3 TPCs and about 120,000 channels overall.
MM modules were produced and tested on a test bench at CERN 
by a complete scan of all the active area with a $^{55}{\rm Fe}$ source 
emitting 5.9 keV photons. 
The energy resolution obtained is 8\% and the uniformity over a module is 
good: 3\% in gain and 6\% in energy resolution.

\begin{figure}[htb]
\begin{center}
\includegraphics[width=.6\textwidth]{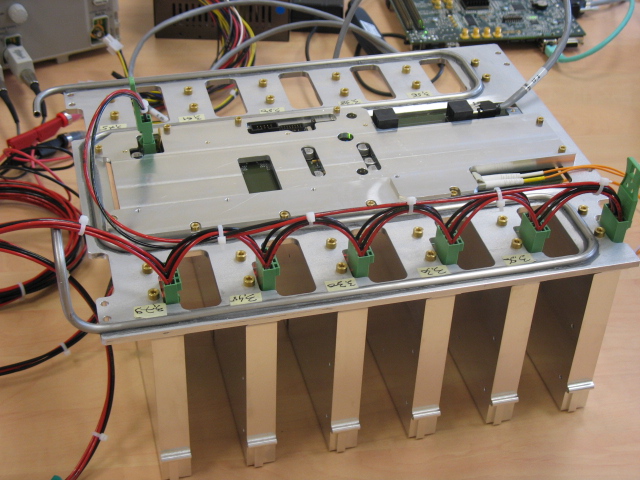}
\end{center}
\caption{View of the readout electronics for one module.}
\label{fig:elec}
\end{figure}

The front-end electronics is based on the asic AFTER. 
This is a chip with 72 channels where the gain, the peaking time, 
and the sampling frequency are programmable. 
They are set to 120 fC, 200 ns, and 25 MHz respectively in T2K data taking. 
There are four asics per front-end card (FEC), which reads out 
one sixth of a module. 
The front-end mezzanine (FEM), one per module, does the data collection 
from six FECs and applies the zero suppression. 
A FEM and its associated six FECs are shown on figure~\ref{fig:elec}.
The data are then sent by optic fibers outside the TPC 
to 18 data concentrator cards (DCC) and then to the data acquisition system.

\begin{figure}[htb]
\begin{center}
\includegraphics[width=.45\textwidth]{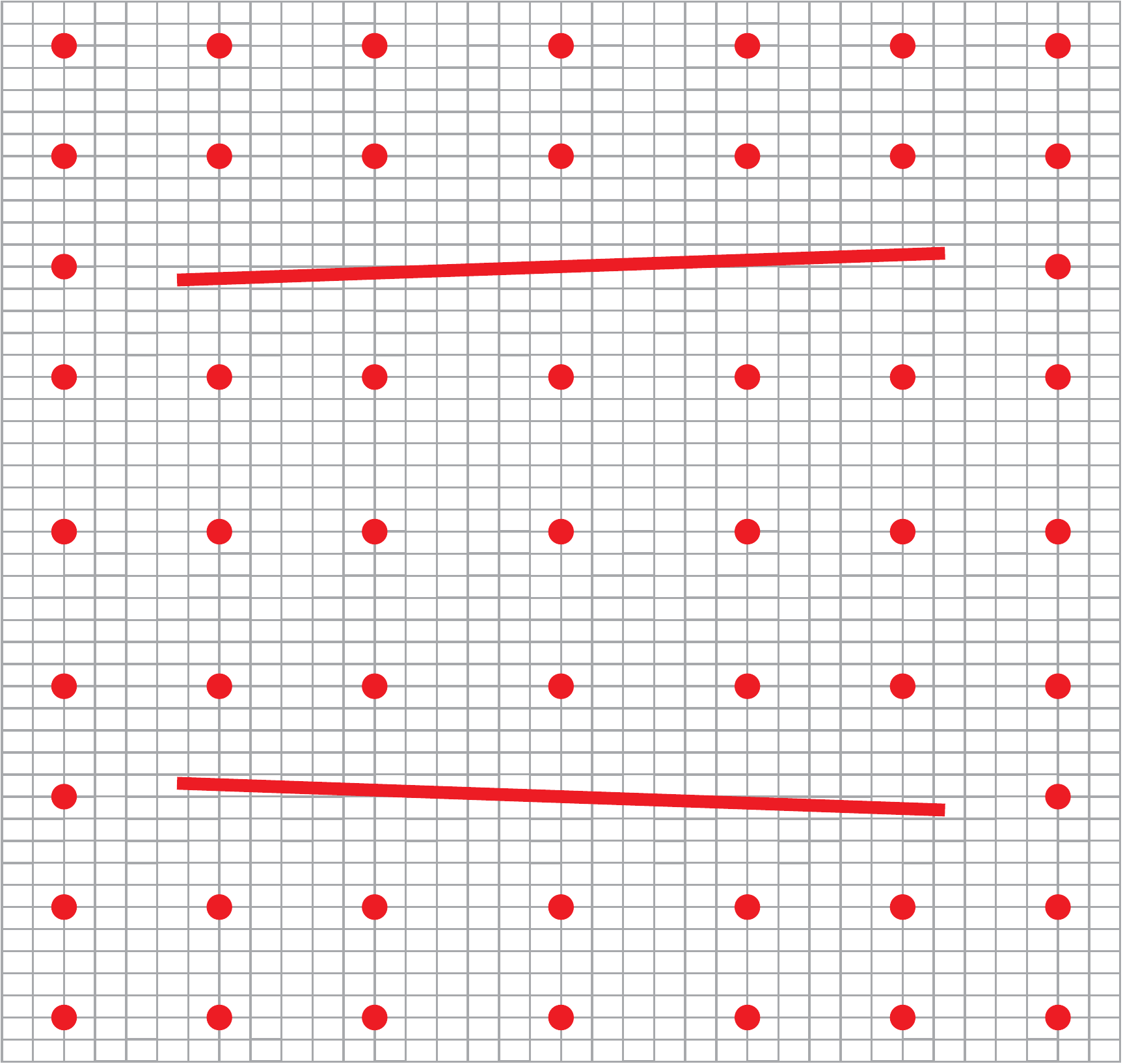}
\end{center}
\caption{Laser target pattern}
\label{fig:laser}
\end{figure}

For calibration purpose, the TPCs have a laser system. 
It emits 266 nm UV light, brought through a set of optical fibers. 
The target pattern, illustrated in figure~\ref{fig:laser},
is made of aluminum discs and strips glued on the cathode, 
which emit photoelectrons when flashed with UV light. 
The laser is used to study electric and magnetic field distortions, 
drift velocity and gain variations.

\section{The time projection chamber performance}

\begin{figure}[htb]
\begin{center}
\includegraphics[width=.6\textwidth]{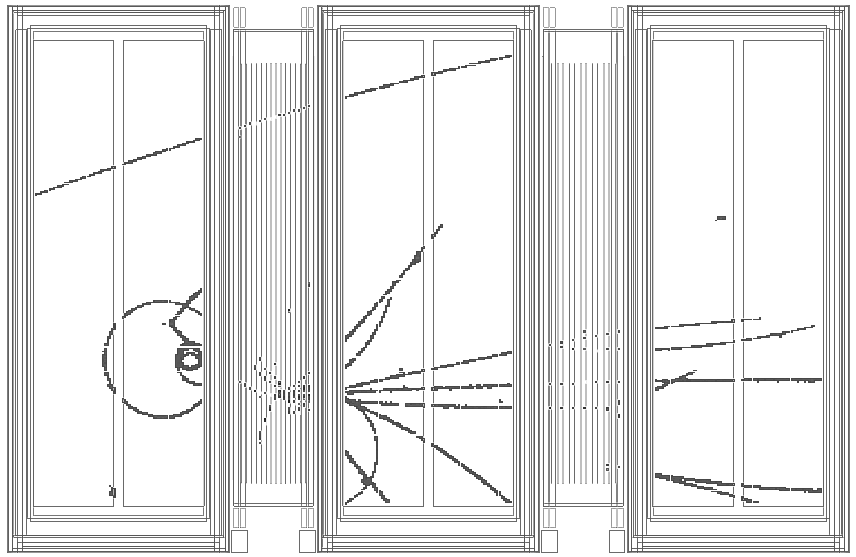}
\end{center}
\caption{Event display showing a deep inelastic scattering neutrino 
interaction in the first FGD (left) with many tracks in the three TPCs. 
There is an additional through going track (top left) 
coming from an upstream neutrino interaction.}
\label{fig:display}
\end{figure}

The production and construction phase took place from 2007 to 2009. 
All three TPCS were assembled and tested with beam at TRIUMF as soon as 2008 
for the first one and in 2009 for the last two. 
Then they were installed at J-PARC and commissioned at the end of 2009. 
They were operational during the first T2K physics run 
from January to June 2010 and the second one from November 2010 to March 2011.
Figure~\ref{fig:display} shows the display of one beam event. 

\begin{figure}[htb]
\begin{center}
\includegraphics[width=.6\textwidth]{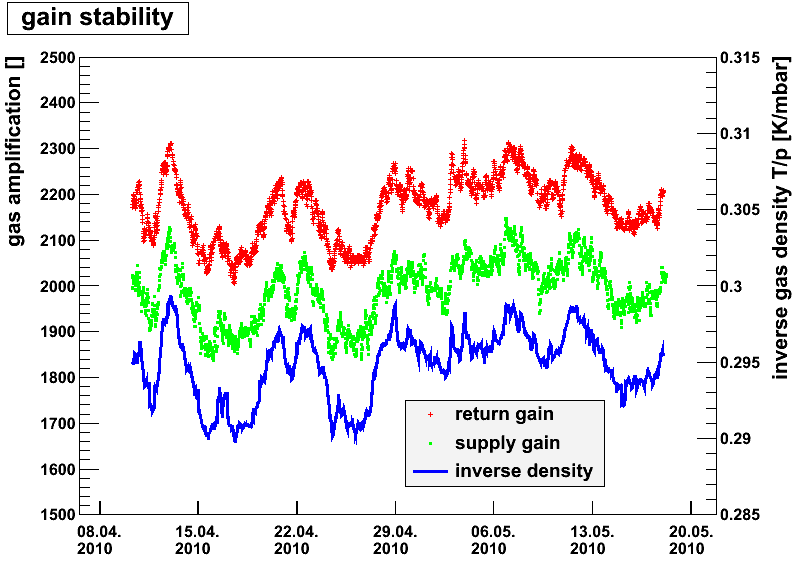}
\end{center}
\caption{Evolution with time of the gain measured in the supply line 
(green curve) and return line (red curve) and of the inverse gas density 
(blue curve).}
\label{fig:gainvar}
\end{figure}

The TPC life fraction during beam time is nearly 100\%.  
MM modules have a spark rate of about 0.1 per hour at 350 V,
inducing negligible dead time. 
The front-end electronics electric consumption is 2.8 kW. 
The rate of TPC data to DAQ is less than 2MB/s with a trigger rate at 20 Hz. 
During data taking, many quantities are monitored including gain, gas density,
gas quality, drift velocity, temperatures, voltages and currents.
For the gain for example, the two monitoring chambers show large variations 
of the gain overtime, shown in figure~\ref{fig:gainvar}.
But these variations are pretty much correlated with the variation 
of the inverse gas density, which is mainly due to changes in temperature and 
pressure. 
As theses quantities are also monitored, 
T and P corrections can be applied to the gain. 
After that, the gain is stable within 1\%.

\begin{figure}[htb]
\begin{center}
\includegraphics[width=.4\textwidth]{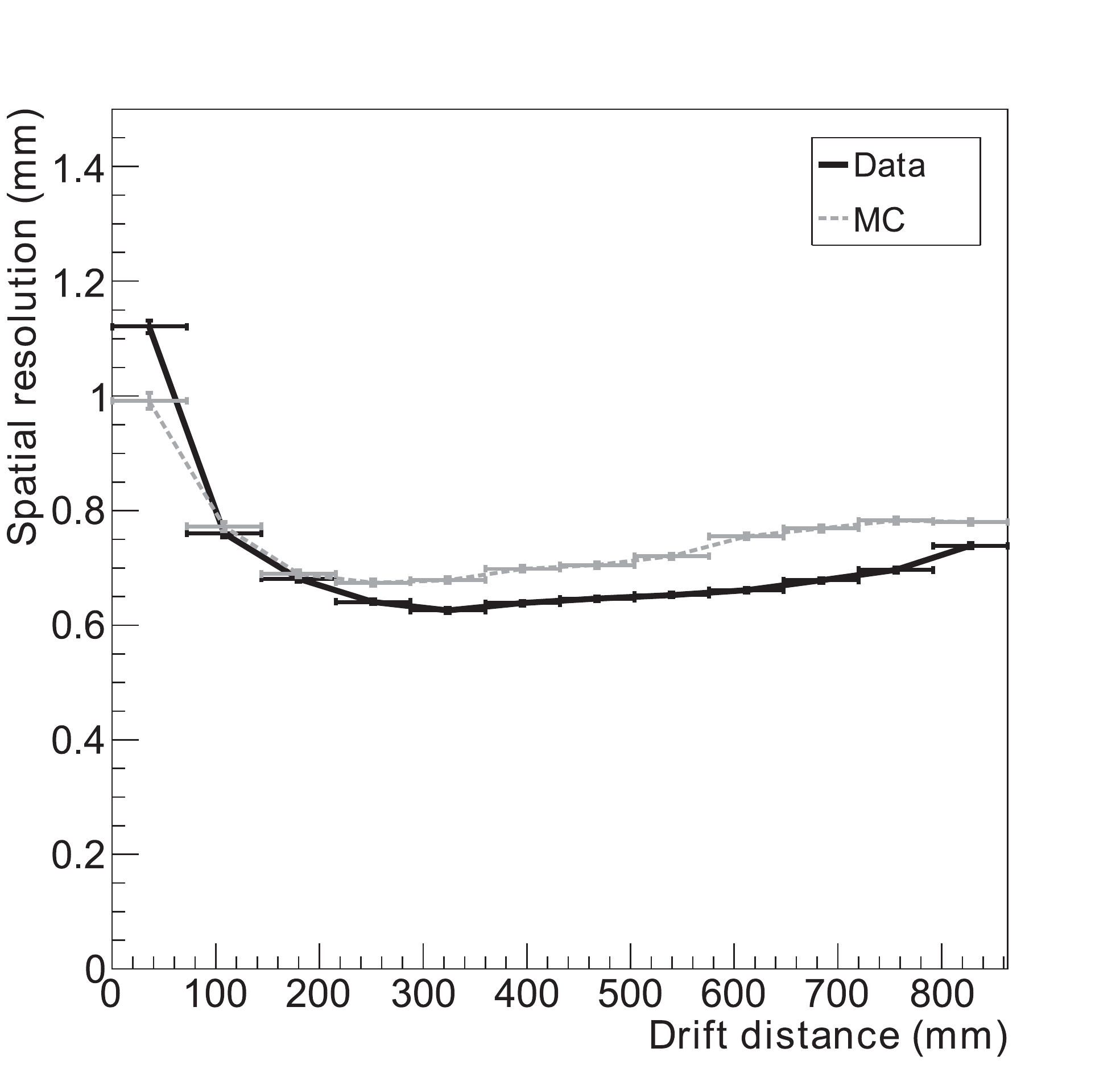}
\includegraphics[width=.4\textwidth]{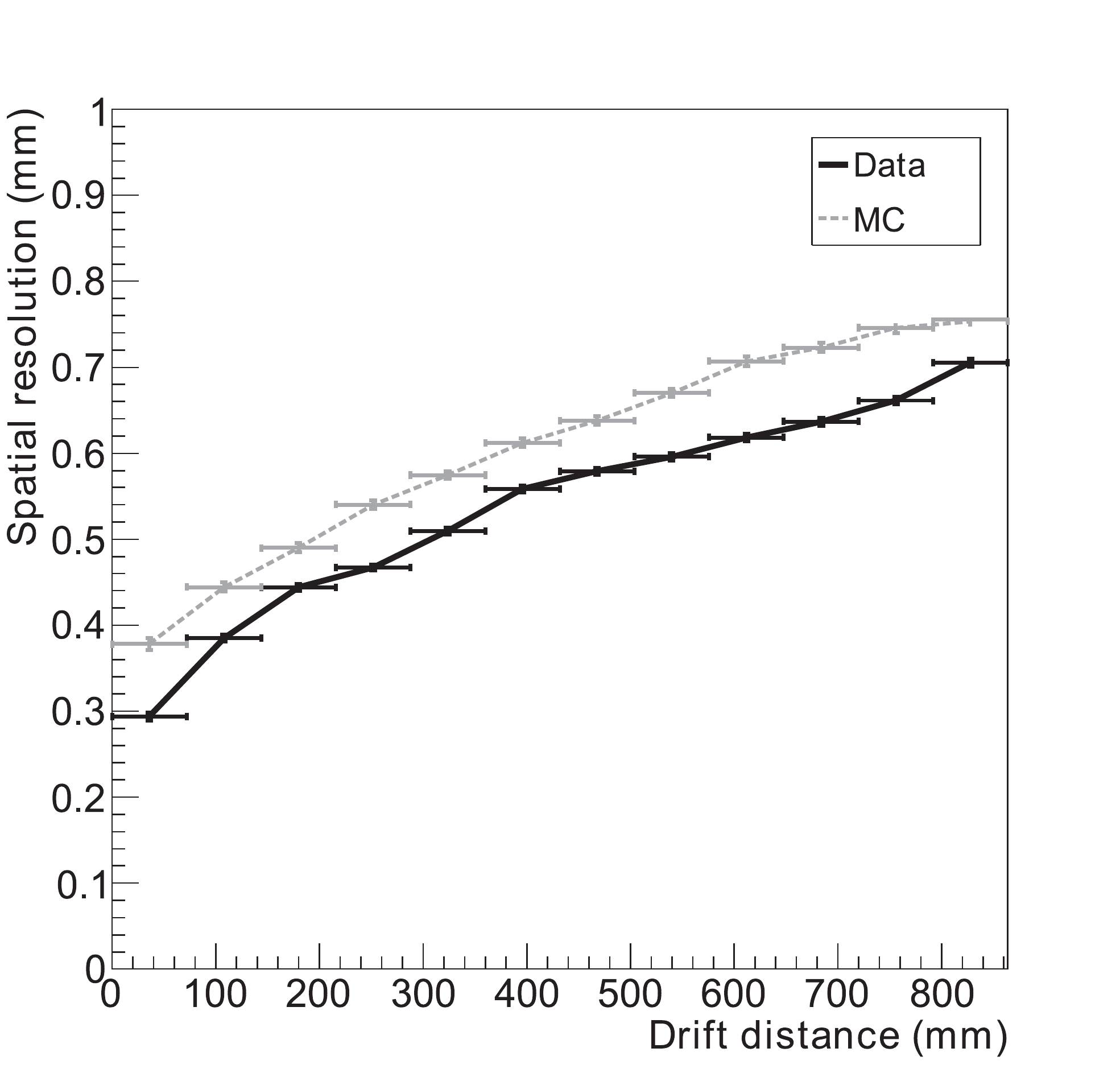}
\includegraphics[width=.4\textwidth]{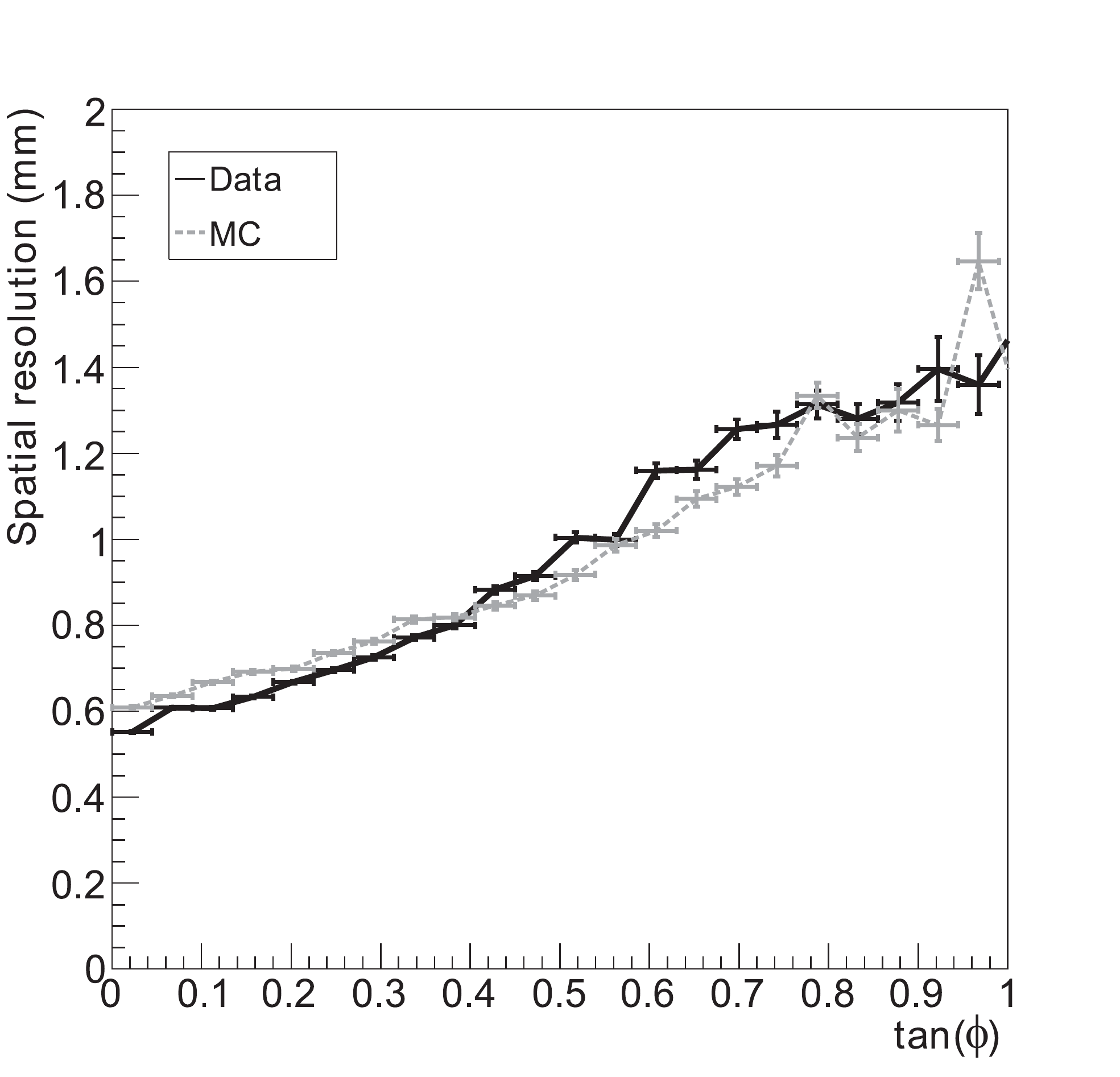}
\end{center}
\caption{Spatial resolution (top) as a function of drift distance for (left)
all clusters and (right) two-pad clusters, and (bottom) as a function of the
tangent of the angle in the vertical plane. 
Data are shown by the black points and solid lines, while simulations are
given by the gray points and dashed lines.}
\label{fig:spatial}
\end{figure}

In the reconstruction, clusters are formed from neighboring pads either 
within a column for horizontal tracks or within a line for vertical tracks. 
A track is a set of such clusters. 
For each TPC, the spatial resolution is estimated by comparing 
the transverse coordinate resulting from the global track fit 
to the one obtained with a single cluster fit. 
As shown in figure~\ref{fig:spatial} (top left), 
it is smaller than 800 $\mu$m for drift length above 150 mm. 
The degraded resolution at short drift distance is due to the larger fraction 
of single pad clusters. 
For two-pad clusters, the increase of the resolution with drift distance seen
in figure~\ref{fig:spatial} (top right) shows the effect of the diffusion. 
Finally the strong dependence of the spatial resolution upon the angle 
in the vertical plane, illustrated in figure~\ref{fig:spatial} (bottom),
 is due the ionization fluctuation along the track. 
We get a spatial resolution of 600 $\mu$m for horizontal tracks.
There is generally a good agreement between the simulated and 
the measured spatial resolution. 

\begin{figure}[htb]
\begin{center}
\includegraphics[width=.4\textwidth]{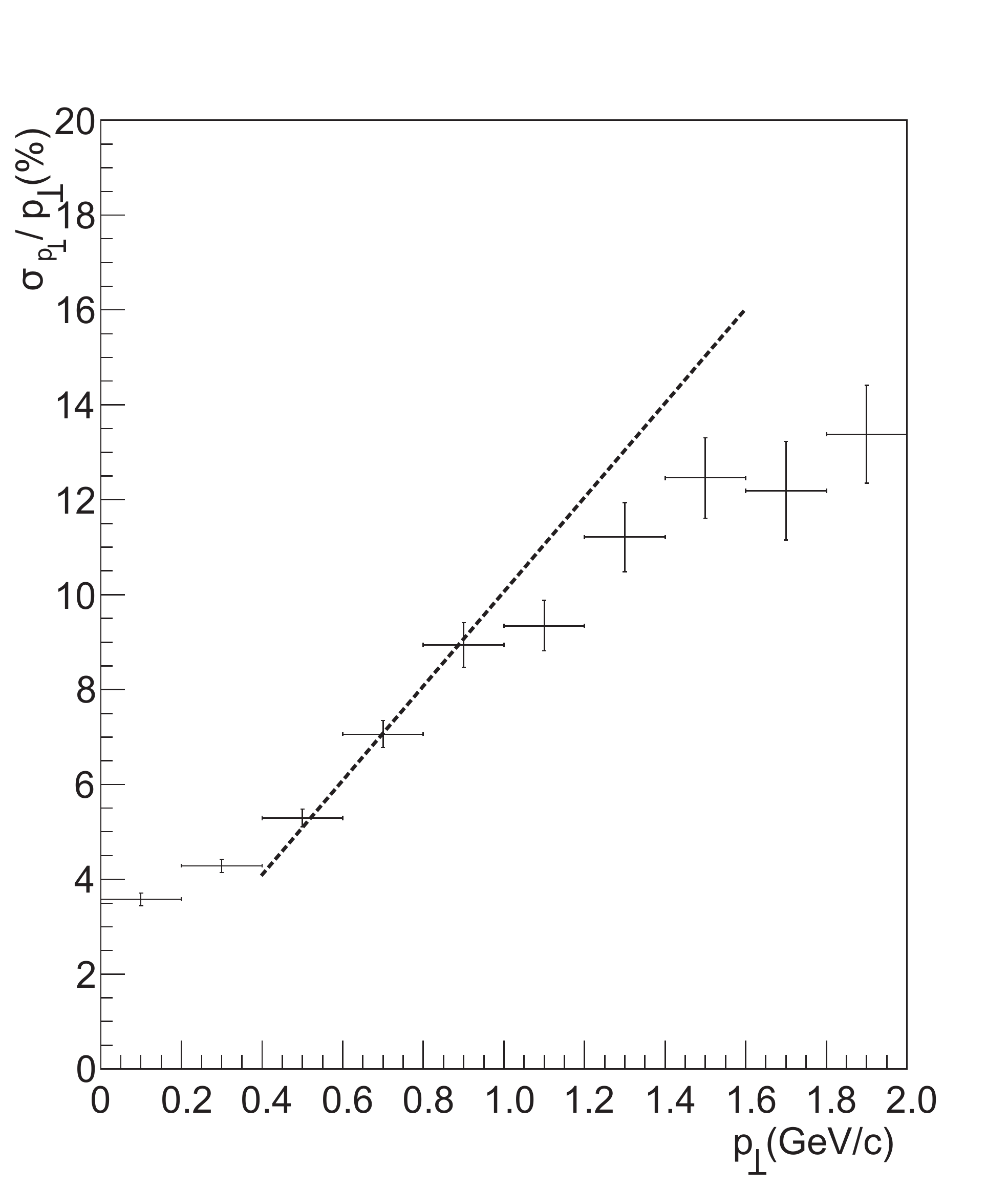}
\includegraphics[width=.5\textwidth]{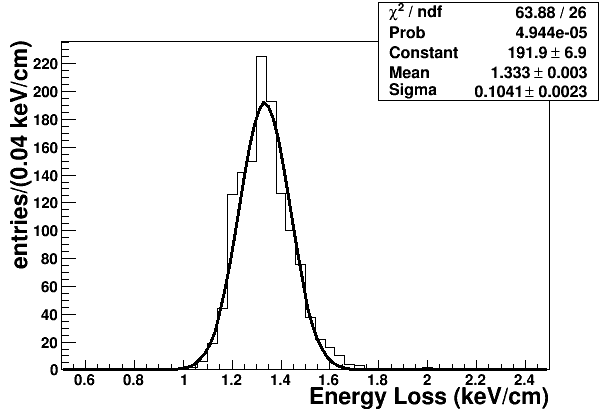}
\end{center}
\caption{Left: relative transverse momentum resolution for a single TPC
as a function of transverse momentum. 
The dashed line shows the required resolution.
Right: distribution of the energy loss for negatively charged particles 
with momenta between 0.4 and 0.5 GeV/c.}\label{fig:momres}
\end{figure}

The relative resolution on the track transverse momentum with respect 
to the B-field direction for a single TPC is estimated with a Monte Carlo 
sample of muons obtained from simulated neutrinos interactions. 
The result illustrated in figure~\ref{fig:momres} (left) shows that 
for transverse  momenta above 0.4 GeV/c the required relative momentum 
resolution of 10\% is achieved. 

Particle identification uses an energy loss truncated mean.
Figure~\ref{fig:momres} (right) shows that a relative resolution on the energy
loss of 7.8\% is obtained, so here again better than the design goal of 10\%.

The distortions of electric or magnetic field, 
which affect the reconstructed track parameters, are studied 
using laser calibration data. 
Electric field distortions are obtained with the magnet off 
from the observed offsets of the aluminum discs on the cathode 
with respect to the geometrical survey. 
The rms values obtained in both directions are smaller 
than the space point resolution. 
As for the magnetic field distortion, they are investigated 
by comparing the measurements done with and without magnetic field. 
The offsets are typically less than 1mm, 
but reaching 5 mm in some regions of the downstream TPC at maximum.  

As a consequence of the earthquake on March 11, 2011,
T2K data taking was stopped. 
The status of ND280 could be checked in details in May 2011. 
In particular, for TPC, the gas system, cooling, the front-end electronics, 
the back-end electronics, the high voltages for cathodes and 
micromegas modules were all powered up successfully. 
So the TPCs were brought back to full operation and 
should be ready to resume data taking when the beam is back. 
The current plane is to restart beam operations in December 2011 and 
start a new T2K physics run in the beginning of 2012.

\section{Conclusion}

In summary, the three T2K ND280 TPCs have been operating successfully 
during the first two physics run from January 2010 to March 2011. 
They will be ready for the next T2K physics run foreseen for the beginning 
of 2012. 
They are contributing in an essential way to T2K physics results
\cite{claudio}.

\end{document}